\documentclass[runningheads]{llncs}
\usepackage{graphicx}
\usepackage{siunitx}
\usepackage{amsmath}
\usepackage{amssymb}

\usepackage{caption}
\usepackage{float}
\usepackage{cite}
\usepackage{xcolor}

\usepackage{algorithm}
\usepackage{algpseudocode}

\DeclareMathOperator*{\argmin}{\arg\min}

\def\denoise{{\mathrm{denoise}}}
\def\prox{{\mathrm{prox}}}
\def\register{{\mathrm{register}}}
\usepackage[normalem]{ulem}

\begin{document}
\title{Plug-and-Play Priors for Reconstruction-based Placental Image Registration}
%

\author{
Jiarui Xing\inst{1} 
\and Ulugbek Kamilov\inst{2,3} 
\and Wenjie Wu\inst{4,5}
\and Yong Wang\inst{5} 
\and Miaomiao Zhang\inst{1,6}}
\authorrunning{J. Xing et al.}
\institute{
Electrical and Computer Engineering, University of Virginia, Charlottesville, USA
\and Computer Science and Engineering, Washington University in St. Louis, USA
\and Electrical and Systems Engineering, Washington University in St. Louis, USA
\and Biomedical Engineering, Washington University in St. Louis, USA.
\and Obstetrics and Gynecology, Washington University in St. Louis, USA.
\and Computer Science, University of Virginia, Charlottesville, USA
}

%
%
%
\maketitle              
%
%
\begin{abstract}
This paper presents a novel deformable registration framework, leveraging an image prior specified through a denoising function, for severely noise-corrupted placental images. Recent work on plug-and-play (PnP) priors has shown the state-of-the-art performance of reconstruction algorithms under such priors in a range of imaging applications. Integration of powerful image denoisers into advanced registration methods provides our model with a flexibility to accommodate datasets that have low signal-to-noise ratios (SNRs). We demonstrate the performance of our method under a wide variety of denoising models in the context of diffeomorphic image registration. Experimental results show that our model substantially improves the accuracy of spatial alignment in applications of 3D in-utero diffusion-weighted MR images (DW-MRI) that suffer from low SNR and large spatial transformations.

\end{abstract}
%
\section{Introduction}
Placental pathology, such as immune cell infiltration and inflammation~\cite{blencowe2013born}, is a common reason for preterm labor. It occurs in around $11 \%$ percent of world pregnancies. \emph{Diffusion-weighted magnetic resonance imaging (DW-MRI)} is a non-invasive technique that is extensively used to monitor placental health and to assess its function throughout the entire pregnancy. However, this method is quite susceptible to motion artifacts caused by maternal breathing and fetal movements~\cite{le2006artifacts}. Additionally, DW-MRI scans often suffer from noise and severe artifacts induced by low signal-to-noise ratios (SNRs) at high b-values~\cite{partridge2013diffusion,vishnevskiy2015simultaneous}. To address these issues, a noise-robust registration algorithm is needed. 

Many attempts have been made to develop registration methods that are robust to image noise~\cite{han2006variational, sanches2003joint, Lombaert2012, lempitsky2007logcut}. A traditional approach integrates an image reconstruction algorithm for removing noise and artifacts as a pre-processing step to the registration task~\cite{tomavzevivc2005reconstruction}. Further improvements can be achieved by developing a joint framework that alternates between image reconstruction and registration~\cite{vishnevskiy2015simultaneous, han2006variational,lombaert2012simultaneous,telea2006variational}. The most widely used image reconstruction algorithms are based on optimization of an objective function that includes a regularization term for mitigating noise. Recently, however, the interest in the area has shifted towards a more flexible approach, known as \emph{plug-and-play priors (PnP)}~\cite{venkatakrishnan2013plug}, that regularizes the problem using off-the-shelf image denoising algorithms. It has been shown that the combination of reconstruction algorithms with advanced denoisers, such as non-local means~\cite{buades2005non} or block matching and 3D filtering (BM3D)~\cite{dabov2007video}, leads to the state-of-the-art performance for various imaging problems~\cite{Sreehari.etal2016, Chan.etal2016, Meinhardt.etal2017, Buzzard.etal2017}.

In this paper, we extend the current family of joint reconstruction-registration algorithms by introducing a new method for deformable image registration called \emph{PnP-RR} (where RR stands for \emph{registration}-\emph{reconstruction}). Our algorithm leverages PnP image priors, which makes it robust for registering severely noise-corrupted images. PnP-RR is very easy to implement by using a wide variety of existing algorithms with minimal effort to modify the infrastructure. We demonstrate how PnP priors can be used to mix and match a wide variety of existing reconstruction models with the state-of-the-art registration algorithm on both 2D synthetic data and real 3D images. To show the effectiveness of the algorithm in improving the performance of spatial alignment for severely noise-corrupted images, we test on 3D in-utero DW-MRI scans, affected by a low signal-to-noise (SNR) ratio and large motions.

\section{Background: Deformable Image Registration}
In this section, we briefly review the mathematical foundation of image registration. Consider a $d$-dimensional image $I$ defined as a continuous mapping $I: \Omega \rightarrow \mathbb{R}^d$, where $\Omega$ is the image domain. The transformation $\phi: \Omega \rightarrow \Omega$ deforms a source image $S$ by function composition $S\circ\phi^{-1}$, where $\circ$ denotes resampling. The goal of image registration is to find an optimal transformation $\phi$, such that the deformed image $S\circ\phi^{-1}$ is similar to a target image $T$. 

The desired transformation $\phi$ is typically computed by minimizing an energy function $E(\phi) = \text{dist}(S\circ\phi^{-1}, T) + \text{reg}(\phi).$
Here, the distance function $\text{dist}(\cdot, \cdot)$ measures the dissimilarity between two images, such as sum-of-squared differences of image intensities\cite{beg2005computing}, mutual information\cite{leventon1997multiple}, and normalized cross correlation\cite{avants2008symmetric}. The regularization term $\text{reg($\cdot$)}$ guarantees the smoothness of the transformation. 
A very original function $\phi$ is defined as a linear function $\phi(x)=x+u(x)$, where $x\in\Omega$ and $u$ is a displacement vector field. With the regularity being set to $\left\| Lu \right \|_{L^2}^2$ ($L$ is a differential operator), the optimization of the energy function $E$ over $u$ arrives at a solution for elastic registration~\cite{broit1981optimal}.

However, such algorithm is not able to avoid geometric artifacts (e.g., folding, tearing, or flipping) of the transformations, especially when large deformation occurs, and may destroy the topology of local structures~\cite{christensen1994deformable}. Instead, an elegant algorithm called large deformation diffeomorphic metric mapping (LDDMM) was developed to ensure a smooth and invertible smooth mapping of $\phi$ between images~\cite{beg2005computing}. The regularization term is defined as an integration over time-dependent velocity fields derived from the transformations. We have the objective function of LDDMM as
\begin{align}
\argmin_{v_t} \frac{1}{\sigma^2}\left\|S\circ \phi^{-1}_1 - T\right\|_{L^2}^2 &+ \int_0^1{(L v_t, v_t) \, \mathrm{d}t}, 
\;\; \text{s.t.} \;\; 
\label{eq:LDDMM-energy}
\frac{\mathrm{d}\phi_t}{\mathrm{d}t} = v_t(\phi_t),
\end{align}
where $\sigma^2$ is a weighting parameter, and $(\cdot, \cdot)$ acts similar to an inner product. 

The optimization of the original LDDMM is solved by gradient-based method over the entire time sequence of $v_t$, which is computationally expensive on high-dimensional images (e.g, a 3D placental MRI with the size of $128^3$). Later, a geodesic shooting algorithm~\cite{miller2006geodesic,vialard2012diffeomorphic} shows that once given an initial velocity $v_0$, the shortest path of $\phi$ can be uniquely determined by integrating the geodesic evolution equation (also known as Euler-Poincare differential equation (EPDiff)) defined by
\begin{equation}
\frac{\mathrm{d}v_t}{\mathrm{d}t} = - K\left[ (D v_t)^T \cdot m_t + D m_t\cdot v_t + m_t\cdot \mathrm{div} \, v_t \right],
\label{eq:EPDiff}
\end{equation}
where $K$ is an inverse operator of the differential operator $L$, $m_t= L v_t$ is a momentum vector living in the dual space of $v_t$, $D$ denotes a Jacobian matrix, and $\mathrm{div}$ is a divergence operator. 

The optimization of Eq.~\eqref{eq:LDDMM-energy} can be equivalently reformulated as
\begin{align}
\label{eq:geodesicshooting}
 \argmin_{v_0} \frac{1}{\sigma^2} \|S\circ \phi^{-1}_1 - T \|_{L^2}^2 + (L v_0, v_0), \, \,  \text{s.t.} \,
 \frac{\mathrm{d}\phi_t}{\mathrm{d}t} = v_t(\phi_t) \, \, \& \, \,  \text{Eq.}~\eqref{eq:EPDiff}.
\end{align}
This effectively shrinks the searching space from a time collection of $\{v_t\}$ to a single initial point $v_0$, thus significantly reducing the computational complexity of the entire optimization. 

It has been recently demonstrated that the initial velocity $v_0$ can be efficiently captured via a discrete low-dimensional bandlimited representation in the Fourier space~\cite{zhang2015finite}. We develop our model by employing this fast registration algorithm named FLASH, which is the start-of-the-art variant of LDDMM with geodesic shooting algorithm~\cite{zhang2017frequency,zhang2019fast}.

\section{Our Method: Image Registration with PnP Priors}
In this section, we introduce a novel noise-robust registration model that incorporates a PnP prior as an additional image regularizer. We show that the our model can be implemented using two independent software modules -- one for image reconstruction and the other for image registration. Therefore, changing the prior model only involves the implementation of image reconstruction. That is to say, our framework can be used to match a wide variety of priors with a suitable registration model. 

\subsection{Formulation as a Proximal Algorithm}
We first consider the following joint objective function that builids on Eq.~\eqref{eq:geodesicshooting} to combine image regularization with deformable registration
\begin{align}
\label{eq:ObjectiveFunc}
 \mathcal{F}(v_0, \tilde{T}) = \frac{1}{\sigma^2} \|S\circ \phi^{-1}_1 - \tilde{T} \|_{L^2}^2 &+ (L v_0, v_0) + \lambda_1 \mathcal{R}(\tilde{T}) + \lambda_2 \|T - \tilde{T} \|_{L^2}^2,
\end{align}
where $T$ is the target image, $\tilde{T}$ is the reconstructed image, $\mathcal{R}(\cdot)$ is the regularization term characterizing the prior on the image, $\lambda_1$ is the parameter controlling the strength of regularization, and $\lambda_2$ controls the fidelity of the reconstructed and noisy images.

In order to solve the problem~\eqref{eq:ObjectiveFunc} efficiently, we adopt an \emph{alternating minimization} approach~\cite{Nocedal.Wright2006}, where $v_0$ is first minimized for a fixed $\tilde{T}$ under the constraints in Eq.~\eqref{eq:geodesicshooting} and vice versa, as follows
\begin{subequations}
\label{Eq:AM}
\begin{align}
\label{Eq:AM1}&v_0^k = \argmin_{v_0} \mathcal{F}(v_0, \tilde{T}^{k-1}), \, \,  \text{s.t.} \,
 \frac{\mathrm{d}\phi_t}{\mathrm{d}t} = v_t(\phi_t) \, \, \text{and} \, \,  \text{Eq.}~\eqref{eq:EPDiff}, \\
\label{Eq:AM2}&\tilde{T}^k = \argmin_{\tilde{T}} \mathcal{F}(v_0^k, \tilde{T}),
\end{align}
\end{subequations}
where $k$ denotes the $k$-th iteration. 

By ignoring the terms independent of $v_0$, the step~\eqref{Eq:AM1} can be expressed as
\begin{align*}
\label{Eq:IterRegis}
v_0^k &= \register_{\sigma}(S, \tilde{T}^{k-1}) =\argmin_{v_0} \frac{1}{\sigma^2} \|S\circ \phi^{-1}_1 - \tilde{T}^{k-1}\|_{L^2}^2 + (L v_0, v_0),
\end{align*}
where we didn't explicitly write the constraints for better readability. Note that this step precisely matches the deformable image registration problem in Eq.~\eqref{eq:geodesicshooting}. Similarly, the step~\eqref{Eq:AM2} can be simplified to the following form

\begin{equation}
\label{Eq:ProxOperator}
\tilde{T}^k = \prox_{\tau \mathcal{R}}(Z^k) =\argmin_{\tilde{T}} \frac{1}{2}\|\tilde{T}-Z^k\|_{L_2}^2 + \tau \mathcal{R}(\tilde{T}),
\end{equation}

where we define
\begin{equation*}
    Z^k = \frac{\lambda_2 T+(1/\sigma^2)(S\circ \phi^{-1})}{\lambda_2 + (1/\sigma^2)} \quad\text{and}\quad \tau = \frac{\lambda_1}{2(\lambda_2 + (1/\sigma^2))}.
\end{equation*}
The minimization problem~\eqref{Eq:ProxOperator} is widely known as the \emph{proximal operator}~\cite{Parikh.Boyd2014} and corresponds to an image denoising formulates as $\mathcal{R}(\cdot)$ regularized optimization. For many popular regularizers, such as $\ell_1$-norm or total variation penalty, the proximal operator either has a closed form solution or can be efficiently implemented~\cite{beck2009fast}, without differentiating $\mathcal{R}(\cdot)$. 

\subsection{Formulation as a PnP Algorithm}
Our alternating minimization algorithm in Eq.~\eqref{Eq:AM} iteratively refines a denoised image $\tilde{T}^k$ by applying the proximal operator  defined in Eq.~\eqref{Eq:ProxOperator}. Recently, the mathematical equivalence of the proximal operator to image denoising has inspired Venkatakrishnan \emph{et al.}~\cite{venkatakrishnan2013plug} to introduce a powerful PnP framework for image reconstruction. The key idea of PnP is to replace the proximal operator in an iterative algorithm with a state-of-the-art image denoiser (e.g., BM3D), which does \emph{not} necessarily have a corresponding regularization function $\mathcal{R}(\cdot)$. This implies that PnP methods generally lose interpretability as optimization problems. Nonetheless, the framework has gained in popularity due to its effectiveness in a range of applications. Additionally, several recent publications have theoretically characterized the convergence and fixed points of PnP algorithms~\cite{Sreehari.etal2016, Chan.etal2016, Buzzard.etal2017, Sun.etal2018a, Ryu.etal2019}. 

Algorithm~\ref{alg:PnPRR} summarizes our PnP-RR algorithm for joint image reconstruction and registration. The fixed point $(v_0^\ast, \tilde{T}^\ast)$ of PnP-RR is defined by a balance between denoising and registration operators, rather than the minimum of a cost function. This makes the algorithm easy to adapt to specific datasets by simply swapping denoisers or registration operators. We corroborate the performance of PnP-RR in the next section by applying it to the challenging problem of image registration under severe amounts of noise.
\begin{algorithm}[h]
\caption{PnP-RR}\label{alg:PnPRR}
\begin{algorithmic}[1]
\State \textbf{input: }Source image $S$, target image $T$, parameters $\lambda_1, \lambda_2$, and $\sigma$
\State \textbf{set: } $\tau = \lambda_1/(2(\lambda_2+(1/\sigma^2)))$
\For{$k = 1, 2, \dots$}
\State $v_0^k \leftarrow \register_\sigma(S, \tilde{T}^{k-1})$\Comment{registration step} 
\State $Z^k \leftarrow (\lambda_2 T + (1/\sigma^2) (S \circ \phi^{-1}))/(\lambda_2 + (1/\sigma^2))$
\State $\tilde{T}^k \leftarrow \denoise_\tau(Z^k)$\Comment{denoising step}
\EndFor\label{euclidendwhile}
\end{algorithmic}
\end{algorithm}








\section{Experimental Evaluation}
To evaluate our proposed method, we test its performance with three existing reconstruction algorithms - total variation (TV)~\cite{rudin1992nonlinear}, total generalized variation (TGV)~\cite{bredies2010total}, and BM3D~\cite{dabov2007video} on both synthetic 2D images and real 3D placental DW-MRI scans with different b-values.

We compare our method with the state-of-the-art fast registration method FLASH~\cite{zhang2019fast} (downloaded from: https://bitbucket.org/FlashC/flashc). In all experiments, we set $L$ as a Laplacian operator, e.g., $L = - (\alpha \triangle + I)^c $ with a positive weight parameter $\alpha = 1.5$ and a smoothness parameter $c = 3.0$. We set $\sigma = 0.015$ and the number of time integration steps $n=10$ across all algorithms. 
We also perform registration-based segmentation and examine the resulting segmentation accuracy of the algorithm. To evaluate volume overlap between the propagated segmentation $A$ and the manual segmentation $B$ for placenta, we compute the Dice Similarity Coefficient $DSC(A, B) = 2(|A| \cap |B|)/(|A| + |B|)$, where $\cap$ denotes an intersection of two regions.

\paragraph{\textbf{Data.}} For 2D synthetic images, we generate a collection of binary images with resolution $100^2$. We then add white Gaussian noise with standard deviation $\sigma = 0.3$ to the target images. 

For real 3D placental DW-MRIs, two healthy pregnant subjects (singleton pregnancies) with gestational age between $20 \pm 1$ weeks were recruited and consented. All subjects were scanned in left lateral position during free breathing. Echo-planar DW-MRIs were acquired on a 3T Siemens VIDA scanner with a $30$ channel phase-array torso coil (FOV $= 386 \times 386 \times 300-330 mm^3$, $3 mm$ isotropic voxels, interleaved slice acquisition, TR = $14600 ms$, TE = $62 ms$, Flip Angle = \ang{90}). Multiple scans with different b values ($b = 0, 75, 100, 150 \, s/mm^2$) were tested and the placenta were manually delineated for images with $b = 0$ by radiologists. All DW-MRIs are of dimension $128 \times 128 \times 50$ and underwent bias field correction, co-registration with affine transformations and intensity normalization.

\paragraph{\textbf{Experiments.}} We first run an experiment on 2D synthetic data registering from a clean source image to a noisy target image, and compare the performance of our method with the baseline algorithm FLASH. For the denoisers, we cross-validate a variety of different parameters and set $\lambda_1=0.045, \lambda_2=0.067$ for TV. Similarly, we have $\lambda_1=0.045, \lambda_2=0.015$ for TGV, and $\lambda_1=0.045, \lambda_2=0.225$ for BM3D. We run each algorithm till convergence.

We run similar experiments on real 3D placental DW-MRIs. MR images with low b-value (e.g., $b=0$) are considered as source images, while others with high b-values (typically noisy images) are target images. After testing a set of different parameters, we set $\lambda_1=0.0225, \lambda_2=0.000225$ for TV, $\lambda_1=0.0338, \lambda_2=0.1$ for TGV, and $\lambda_1=0.0225, \lambda_2=0.1$ for BM3D. To further evaluate the registration accuracy, we measure the Dice score by applying the estimated transformation on manually labeled segmentations of placenta.

\begin{figure}[H]
    \centering
    \includegraphics[width=.7\linewidth]{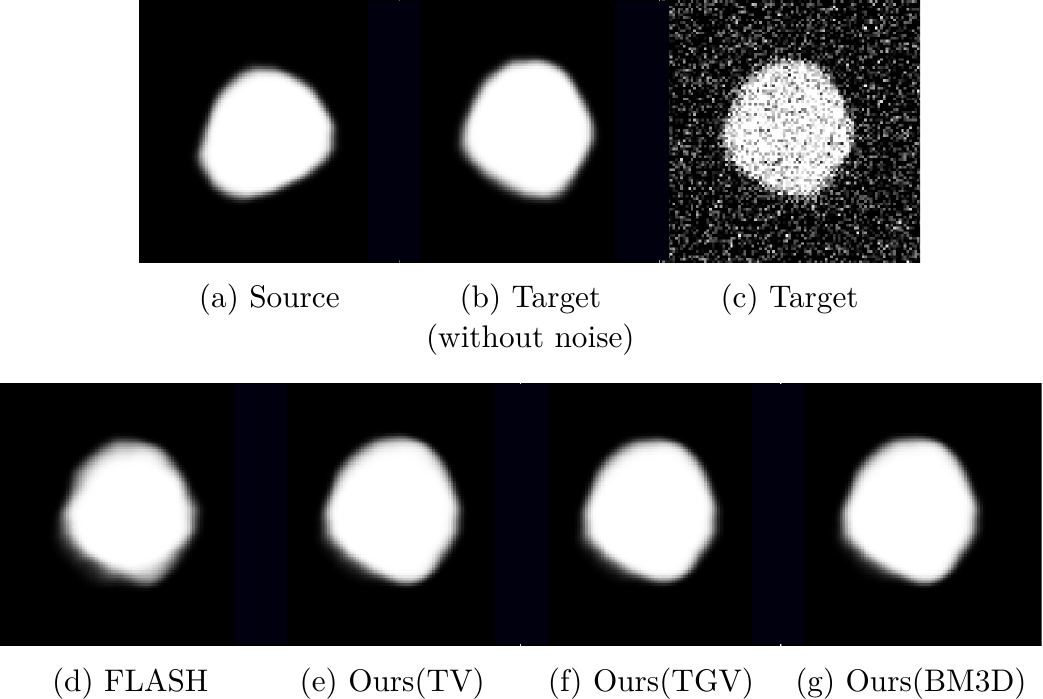}
    \caption{Top: source image, clean target image, and noisy target image; Bottom: registration results from the baseline method FLASH and our model with TV, TGV, and BM3D denoisers.}
    \label{fig:exp-2d-clrcle-img}
\end{figure}

\paragraph{\textbf{Results.}} Fig.~\ref{fig:exp-2d-clrcle-img} displays the registration results of the baseline algorithm and our model with different denoisers. It shows that our method achieves better transformations that nicely deform the source image fairly close to the target image, without being affected by the noises.  

\begin{figure}[H]
    \centering
    \includegraphics[width=0.8\linewidth]{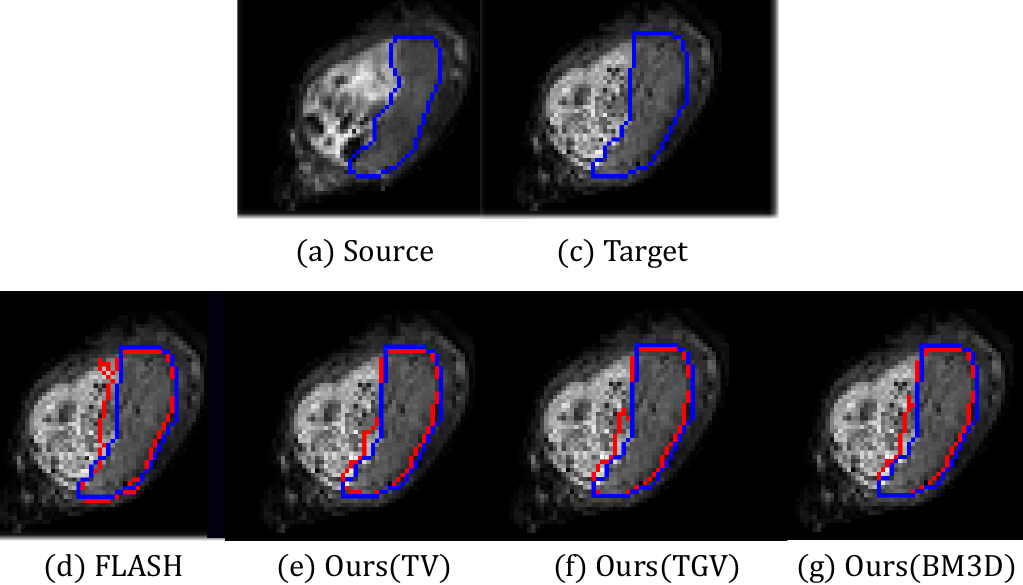}
    \caption{Top: source and target images; Bottom: comparison of estimate segmentations of all algorithms overlapped with manually labeled delineation.}
    \label{fig:exp-placenta-img}
\end{figure}

Fig.~\ref{fig:exp-placenta-img} demonstrates an example of the transformed segmentation of placenta (outlined in magenta) estimated by all algorithms. It clearly shows that the segmentations produced by our algorithm align better with the manual segmentation (outlined in blue) than the baseline algorithm. Our model provides much reliable segmentation than the baseline algorithm, especially on the left part of the placenta where relatively large deformation occurs.

\begin{figure}
    \centering
    \includegraphics[width=.9\linewidth]{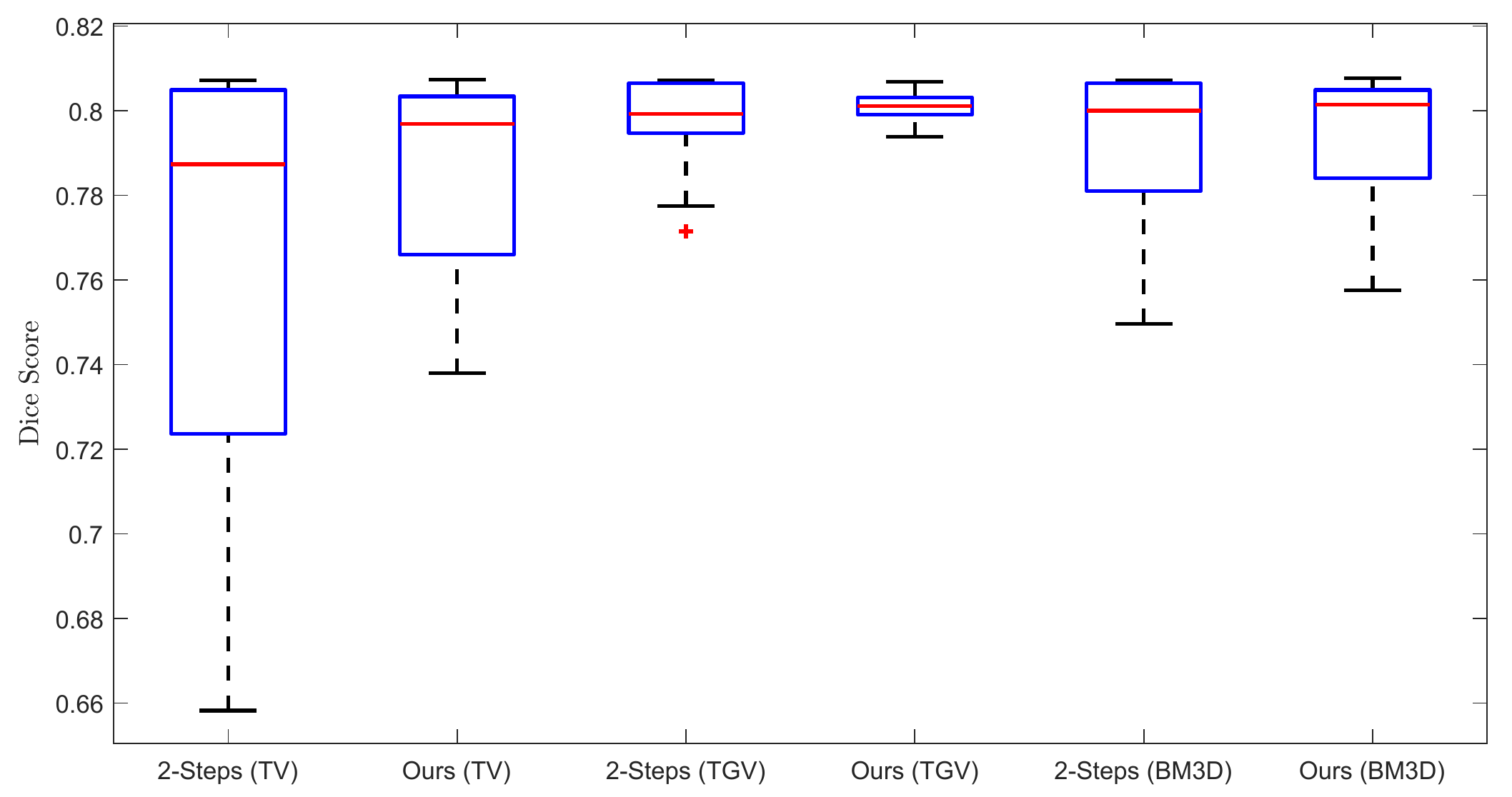}
    \caption{Comparison of averaged Dice score estimated from two-step approaches and ours.}
    \label{fig:dice range}
\end{figure}

Fig.~\ref{fig:dice range} shows another advantage of our model compared to two-step approaches where image reconstruction is preformed before registration. We compute average dice scores with different parameter settings on both methods. Our higher average dice scores with smaller variations indicate that the proposed algorithm is more robust to parameter-tuning. 


\section{Conclusion}
In this paper, we presented a novel reconstruction-based registration algorithm, named PnP-RR, for severely noise-corrupted images. Our method is the first to introduce PnP priors, represented through denoising functions, into the state-of-the-art registration framework. In contrast to previous approaches, our model has the flexibility to allow any reconstruction algorithm integrated with the registration task. This provides a much more robust way to register images with low SNRs and large motions. The theoretical tools developed in our work are broadly applicable to a wide variety of joint reconstruction-registration algorithms. In addition, our method can be easily implemented through the current implementation of registration and reconstruction algorithms. Future research will involve collecting more dataset on placental images and exploring other cutting-edge denoisers, such as deep learning based approaches.

\paragraph{\bf Acknowledgement.} This work was supported by NIH grant R01HD094381, NIH grant R01AG053548, and BrightFocus Foundation A2017330S.

%
%
%
\bibliography{sections/references.bib}
\bibliographystyle{splncs04}
\end{document}